\documentclass[twocolumn,superscriptaddress, english, a4paper]{revtex4}
\usepackage[latin9]{inputenc}
\usepackage{amsmath}
\usepackage{amssymb}
\usepackage{graphicx}
\usepackage{esint}
\usepackage{color}

\makeatletter

\usepackage{babel}

\makeatother

\begin{document}

\title{Leaky Fermi accelerators}

\author{Kushal Shah}
\email{kkshah@ee.iitd.ac.in}
\affiliation{Dept of Electrical Engineering, Indian Institute of Technology (IIT)
Delhi, New Delhi 110016, India.}

\author{Vassili Gelfreich}
\email{v.gelfreich@warwick.ac.uk}
\affiliation{Mathematics Institute, University of Warwick, Coventry CV4 7AL, United Kingdom.}

\author{Vered Rom-Kedar}
\email{Vered.Rom-Kedar@weizmann.ac.il}
\affiliation{Dept of Computer Science and Applied Mathematics, Weizmann
Institute of Science, Rehovot 76100, Israel.}

\author{Dmitry Turaev}
\email{dturaev@imperial.ac.uk}
\affiliation{Dept of Mathematics, Imperial College, London SW7 2AZ, United Kingdom,\\ Lobachevsky University of Nizhny Novgorod, 603950 Russia.}

\begin{abstract}
A Fermi accelerator is a billiard with oscillating walls. A leaky accelerator interacts with an  environment of an ideal gas at equilibrium  by exchange of particles through a small hole on its boundary. Such interaction may heat the gas: we estimate the net energy flow through the hole under the assumption that the particles inside the billiard do not collide with each other and remain in the accelerator for sufficiently long time.  The heat production is found to depend strongly on the type of the Fermi accelerator. An ergodic accelerator, i.e. one which has a single ergodic component,
produces a weaker energy flow than a multi-component accelerator.
Specifically, in the ergodic case the energy gain is independent of the hole size, whereas in the multi-component case the energy flow may be significantly increased by shrinking the hole size. 
\end{abstract}
\maketitle

\section{Introduction\label{sec:Introduction}}
\begin{figure*}
\includegraphics[width=\textwidth]{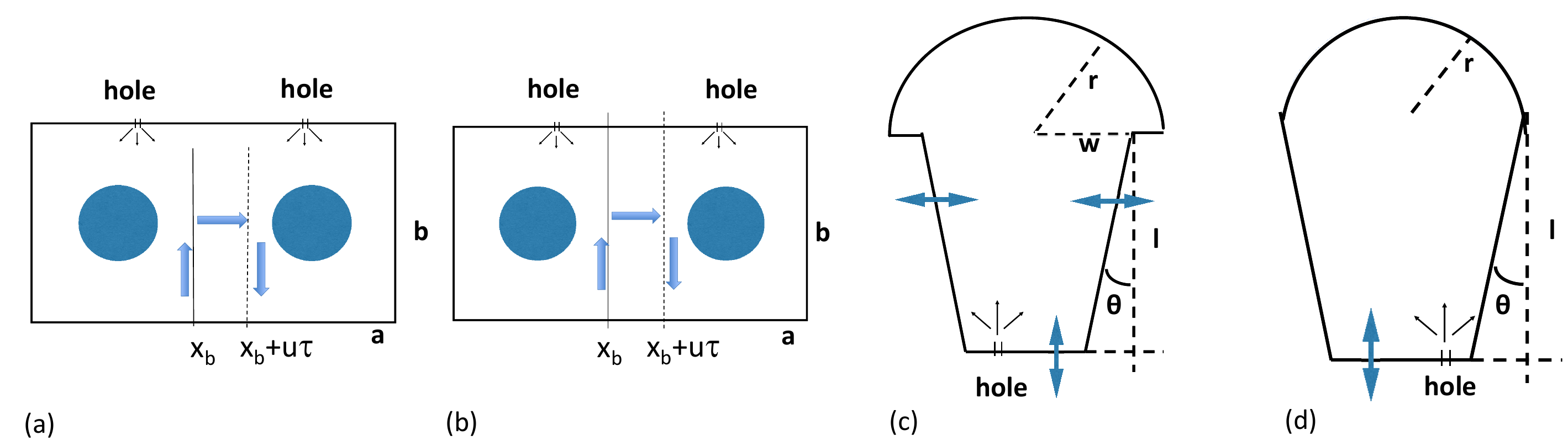}
\caption{(a) Sinai accelerator; (b) divided Sinai accelerator; (c) mushroom accelerator; (d) stadium accelerator. See  simulation section (Sec. \ref{sec:Simulation-Results}) for geometric specifications and dynamical properties.
\label{fig:fig1}}
\end{figure*}

The dynamics of a point particle moving within a closed region (billiard) with oscillating walls provide a mathematical model for studying the phenomenon of Fermi acceleration \cite{Licht, Jarz, Pust, Lieberman, LRA99, Do, GRT}. Such systems typically produce an increase in the particle's kinetic energy and much effort is devoted to quantify this phenomenon. It has been shown that collisions with periodically oscillating walls of an ergodic chaotic billiard accelerate the particle so that
on average its energy grows
linearly with the number of collisions and quadratically as a function of time \cite{LRA99,LRA,GRST, GRT}.
In particular, this behaviour is observed in a periodically oscillating dispersive billiard  \cite{LRA99,GRST}
and in a stadium with  oscillating base  \cite{LRA}.
Examples of such billiards are shown in Figs.~\ref{fig:fig1} (a) and (d).
It was discovered in \cite{STR,GRST} that if the ergodicity of the frozen billiard is violated, i.e. the shape of the billiard is changed in such a way that
several ergodic components are created during a part of the billiard oscillation cycle, then the average energy growth is much faster, typically exponential in time.
A Bunimovich mushroom deformed so as there exists particle exchange between its integrable and chaotic components corresponds to such multi-component, exponential accelerator \cite{GRT14}, see Fig. \ref{fig:fig1}c. The mushroom is a special case of a large class of billiards with mixed phase space where chaotic zones coexist
with stability islands; the exponential character of acceleration at a periodic perturbation of such systems was established in \cite{Benjamin, TD}.
The multi-component accelerators can also be created by pseudo-integrability \cite{Shah, STR} and by division of the billiard configuration space into
disjoint pieces \cite{GRST}, see Fig. \ref{fig:fig1}b.

One of the primary applications of the Fermi acceleration model is in plasma physics where it is used to study the heating of charged particles due to electromagnetic waves \cite{Lieberman, Viana}. In such systems, the electrons absorb energy from the wave in the plasma sheath, deposit it in the plasma bulk and return back to the sheath. Thus, the system is not closed and allows for entry and exit of particles. Leaky chaotic systems also
emerge in numerous physical situations such as chemical reactions, optical microcavities \cite{Yan, Harayama} and hydrodynamic flows (see recent review and references therein \cite{Altmann}).

Stationary leaky billiards have been extensively studied. It was shown that
the escape rate through holes in the billiard boundary depends sensitively on the hole position, size and billiard properties \cite{Altmann,LoMar,CMT,DWY,BY,Bauer}. In this paper we consider time-dependent leaky billiards and provide estimates for energy gain for the two above mentioned classes of the accelerators:
ergodic (like in Figs. \ref{fig:fig1}a,c) and multi-component (Figs. \ref{fig:fig1}b,d). We consider the small hole size limit and observe that $\bar N$, the averaged number of collisions a particle spends in the leaky accelerator, is inversely proportional to the hole size for both cases. 
Then we
demonstrate that for the ergodic case the  averaged energy gain per particle grows linearly with $\bar N$ whereas in the multi-component case the averaged energy gain is much larger and
is approximated by a quadratic polynomial in $\bar N$, see Eqs. (\ref{eq:egainerg}),(\ref{7}) vs. (\ref{eq:egainmc}),(\ref{18}) and Fig. \ref{fig:fig3}.

\begin{figure*}
\includegraphics[width=\textwidth]{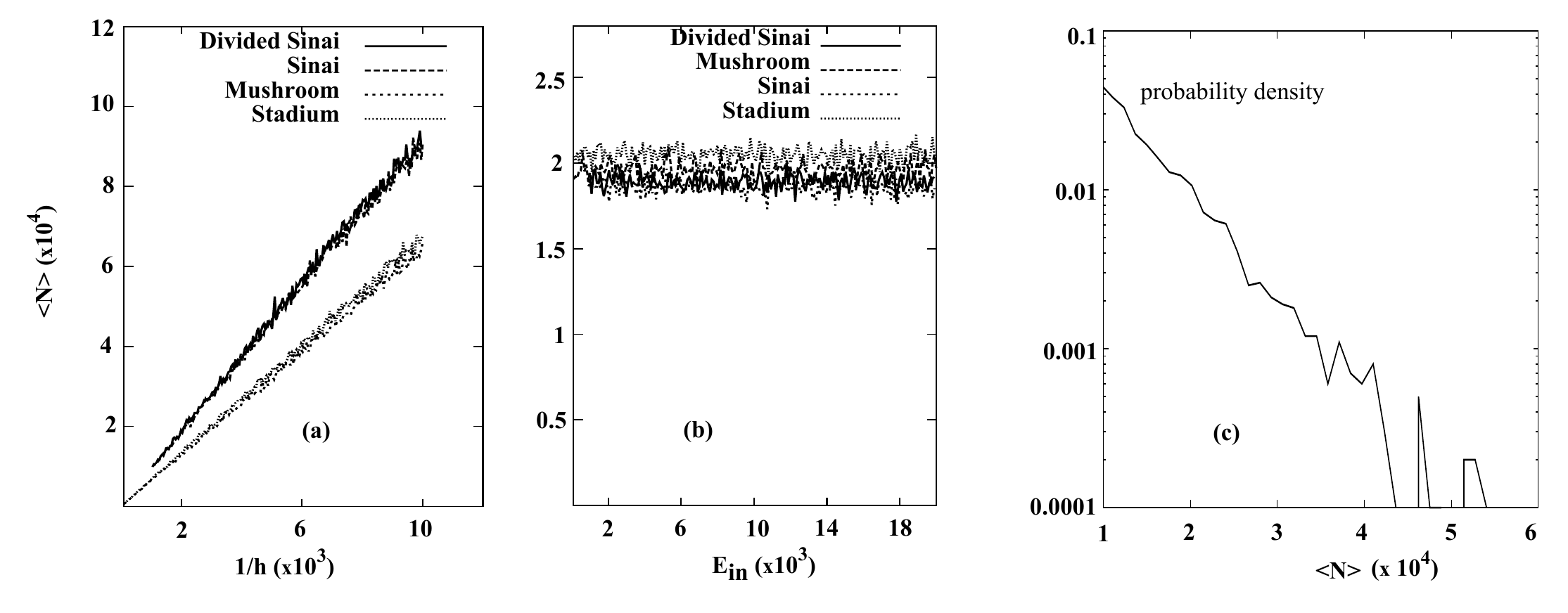}
\caption{The distribution and the average number of collisions, see Eq.~\eqref{5}. (a) dependence on the hole size. Here, the initial energy is $E_{in}=9000$ for the Sinai and divided Sinai accelerators, and  $E_{in}=1250$ for the mushroom and stadium accelerators. (b) no dependence on the initial particles' energy. Here $h=0.0005$ for Sinai and divided Sinai accelerators, and $h=0.00033$ for mushroom and stadium accelerators. (c) the exponential distribution of the exit probabilities for the stadium accelerator. Here $E_{in}=1250$ and $h=0.00033$. \label{fig:fig2}}
\end{figure*}

\begin{figure}
\includegraphics[scale=0.32]{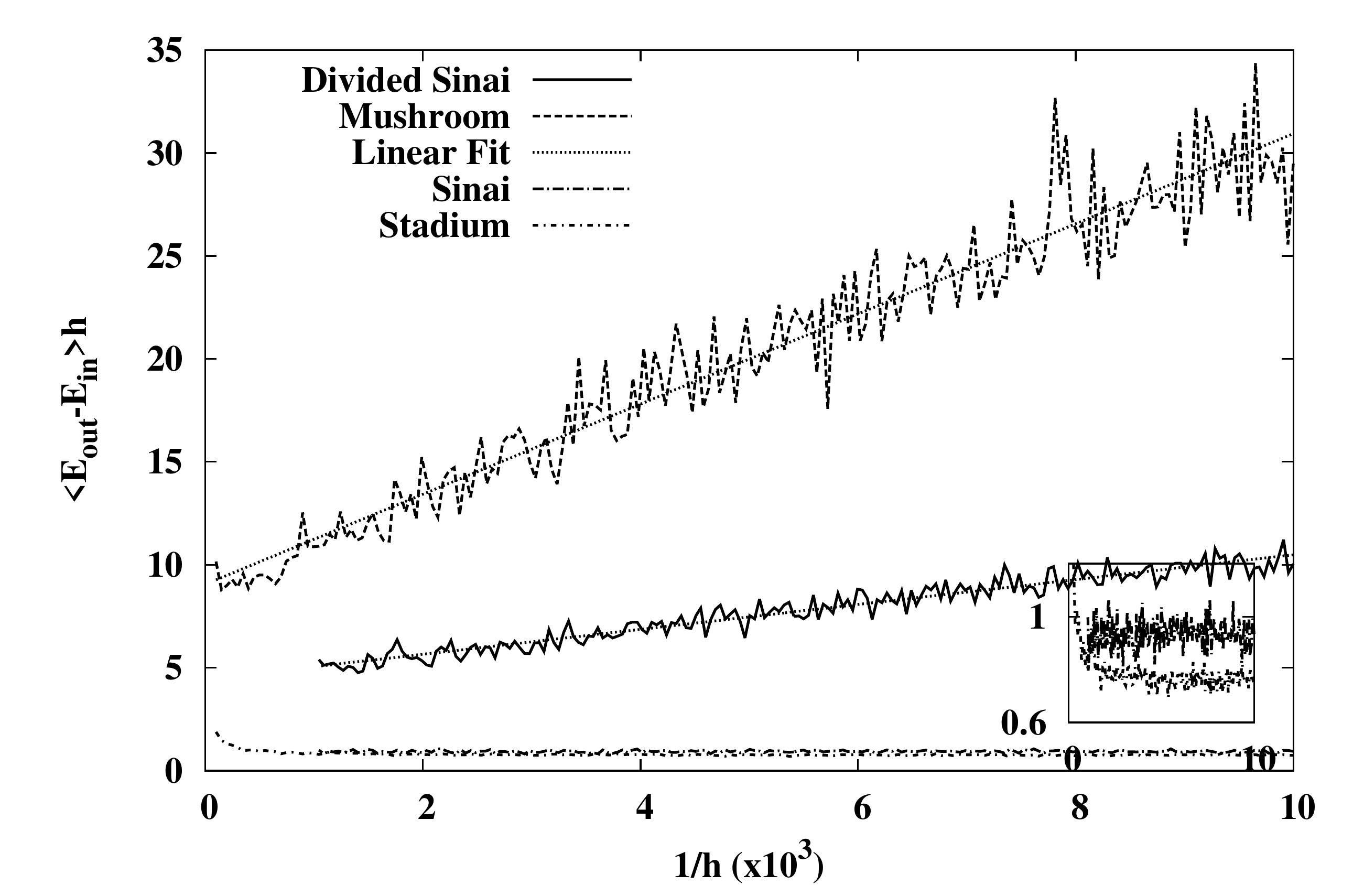}
\caption{Averaged energy gain dependence on the hole size. The energy gain increases linearly with $1\big/h$ for multi-component billiards, see  Eq.~\eqref{18}, and is independent of the hole size for the ergodic billiards (for sufficiently small holes), see Eq.~\eqref{7}. Here, the initial energy is $E_{in}=9000$ for the Sinai and divided Sinai accelerators, and $E_{in}=1250$ for the mushroom and stadium accelerators. \label{fig:fig3}}
\end{figure}

\begin{figure}
\includegraphics[scale=0.32]{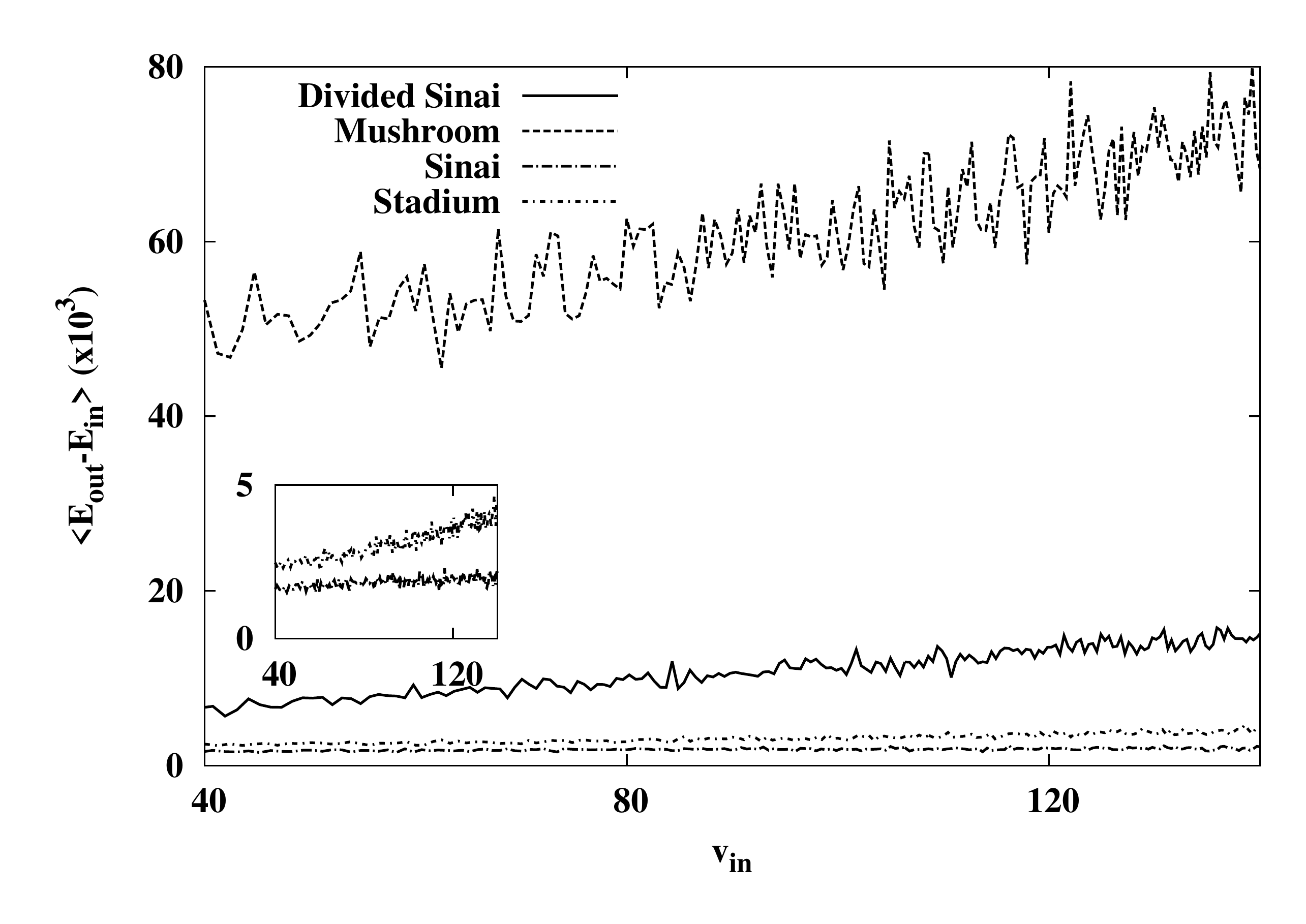}
\caption{Dependence of the average energy gain on the initial speed: linear growth with $v_{in}$ for multi-component billiards,  Eq.~\ref{eq:egainmc}, and slow growth with $v_{in}$ for the ergodic case, possibly due to order \(h\) corrections to Eq.~\ref{eq:egainerg}.
Here $h=0.0005$ for Sinai and divided Sinai accelerators, and $h=0.00033$ for mushroom and stadium accelerators.
\label{fig:fig4}}
\end{figure}

\begin{figure}
\includegraphics[scale=0.32]{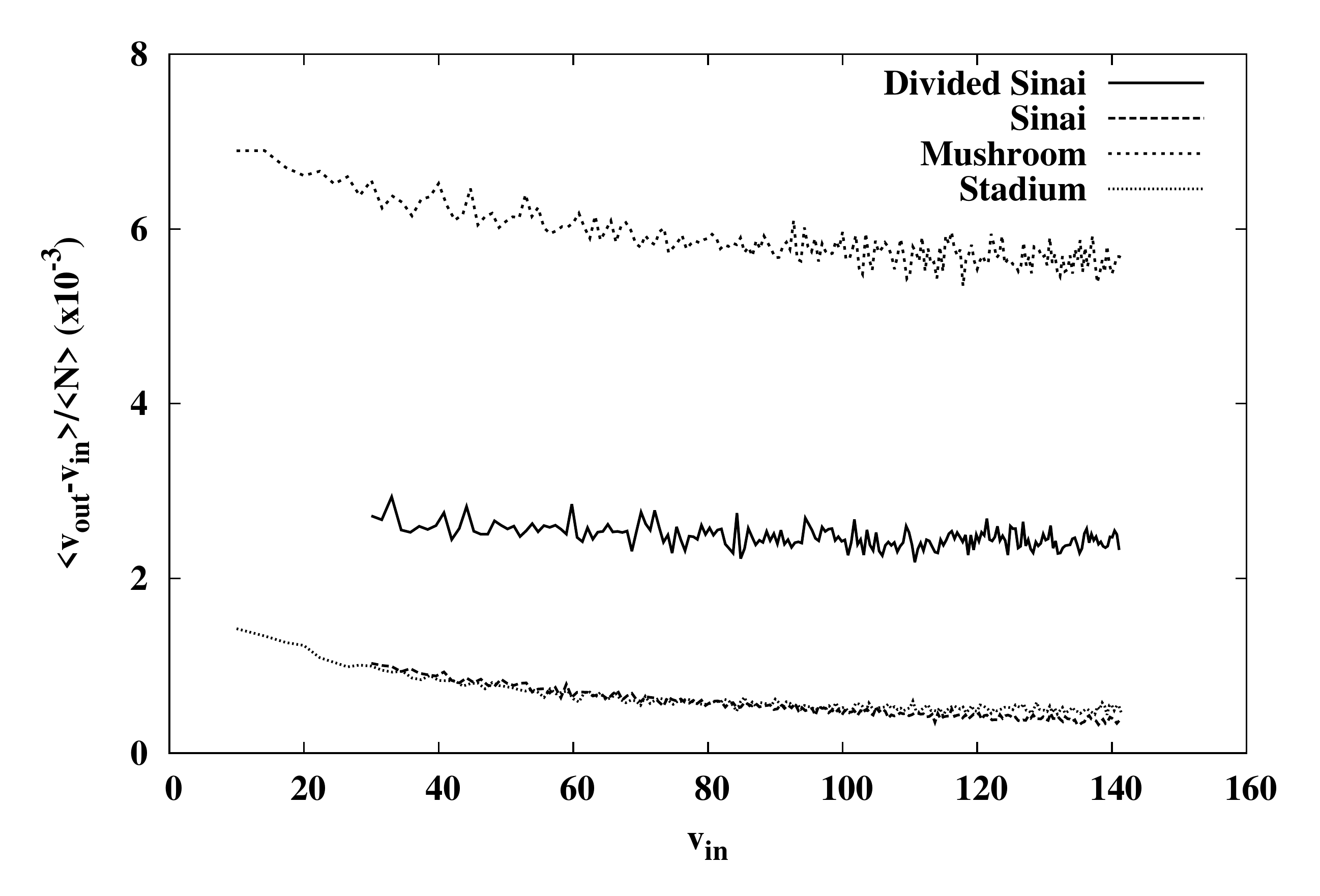}
\caption{The speed bias vs. initial speed. Here $h=0.0005$ for Sinai and divided Sinai, and $h=0.00033$ for mushroom and stadium. \label{fig:fig5}}
\end{figure}

\section{Model\label{sec:Theory}}

Consider an accelerator which interacts with
an ideal gas by exchange of particles through a small hole (or a few small
holes) on its boundary. We assume that the gas is at equilibrium, i.e. there is
a stationary distribution of the particles' speed and the distribution of the angles
at which the particles move in the gas is uniform. We assume that
the particles move much faster than the billiard boundary.
Collisions with the moving billiard walls change the kinetic energy of the particles inside the billiard,
and, on average, this may lead to an outgoing energy gain.

For a fixed kinetic energy $E_{in}=\frac{v_{in}^{2}}{2}$ (we assume
the particles have a unit mass), the number density
of particles entering the billiard per unit of time is proportional to $hv_{in}$,
where $h$ is the size of the hole (the area of the hole for the three-dimensional
case). Thus, the incoming energy flow at energy $E_{in}$ is proportional to $hv_{in} E_{in}$.
We assume that inside the billiard the particles do not collide/interact
with each other, so we can consider each of them separately. This gives us the net energy production
by the accelerator per unit of time:
\begin{equation}
G(E_{in})=hv_{in}\left[E_{out}-E_{in}\right],\label{1}
\end{equation}
where $E_{out}$ is the averaged value of the kinetic energy at the moment of exit for a particle that enters
the accelerator with the energy $E_{in}$ (we average over all possible initial angles and positions in the hole, as
well as over the phase of the billiard oscillations at the entry moment).

Let $p_{_N}$ denote the probability to exit the accelerator after $N$ collisions with the billiards walls
and $\bar{E}(N)$ be the corresponding averaged exit energy. Then
\begin{equation}
E_{out}=\sum \bar{E}(N; E_{in}) p_{_N} \label{eq:eout}
\end{equation}
We assume that the hole size $h$ is small enough, so the effect of the hole on the statistics
of the billiard is negligibly small (as in the case of the stationary Lorentz gas \cite{DWY}). Specifically, we assume that
$\bar{E}(N;E_{in})$ can be approximated by the averaged energy of a particle in the closed (i.e. non-leaky) accelerator after $N$ collisions.
Additionally, we assume that $p_{_N}$ and, thus, $\bar N$, the averaged number of collisions before exit, do not dependent on $E_{in}$
nor on the wall velocity $u$. This is obviously
true when the billiard walls 
are stationary,
so we extrapolate this claim
to the case of slowly moving boundaries. We confirm this claim numerically for the examples we consider here (see Fig. \ref{fig:fig2}).
In fact, the numerics show that $p_{_N}$ can be well approximated by the geometric distribution
$p_{_N}=\frac{1}{\bar N}(1-(1/\bar N))^{N-1}$  (Fig. \ref{fig:fig2}c). The average value $\bar N$ in this setting
is just a geometric characteristic of the billiard and the hole. The natural assumption is
\begin{equation}
\bar{N}\sim\frac{S}{h}\sim\frac{V}{Lh}, \qquad \overline{N^2}\sim (\frac{V}{Lh})^2,\label{5}
\end{equation}
where $h$ is the size of the hole in the billiard boundary, $S$
is the size of the entire billiard boundary, $V$ is the volume occupied
by the billiard, and $L$ is the characteristic diameter of the billiard. Relations \eqref{5}
are confirmed by numerical experiments (see Fig. \ref{fig:fig2}a).

To find the net energy production \eqref{1}, it remains to estimate the dependence of the averaged energy $\langle E \rangle $ of a particle in the closed accelerator on
the number of collisions $N$. Let us recall how energy is gained in the accelerators. The reflection law for a particle hitting a moving wall is obtained by
going to a coordinate frame that moves with the same velocity as the wall
at the moment of collision. In the moving coordinates, we have an elastic reflection law which, after returning to the stationary frame,
results in the reflection law
\begin{equation}
v_{_\bot} '=2u(t,x)-v_{_\bot}, \qquad  v_{_\parallel}^\prime= v_{_\parallel}
\label{eq:reflection}\end{equation}
where $u(t,x)$ is the normal velocity of the wall at the collision point $x$ at the moment $t$;  $v$ and $v^\prime$ are the velocities before and after the collision, and the 
subscripts
$\bot$ and $\parallel$ stand for the components of the velocity which are normal and, resp.,
parallel to the wall.

If the billiard is chaotic, then the correlations
between the consecutive angles $\phi$ at which the particle hits the wall decay fast. Therefore,
the process described by Eq. \eqref{eq:reflection} may be approximated by a random walk with reflections: at each collision the particle velocity
undergoes a reflection and acquires an increment at a random direction.

This random walk proceeds differently for the two main classes of accelerators, ergodic and multi-component \cite{GRT}.
In the ergodic case the random walk becomes unbiased in the large speed limit, which means that the square of velocity
(i.e. the kinetic energy) grows linearly with the number of collisions. Indeed, by taking the square of Eq. (\ref{eq:reflection}),
the energy $E_{_N}=\frac{1}{2} v_{_N}^2$ after the $N$-th collision satisfies
\begin{equation}
E_{_{N+1}}=E_{_N}-2u(t_{_N},x_{_N}) v_{_N} \cos\phi_{_N} +2u^2(t_{_N},x_{_N}).
\label{enrc}
\end{equation}
Since $|u|\ll v$, the change in the billiard shape and relative change in the energy are not significant
for a large number of consecutive collisions, so one can average Eq. \eqref{enrc} over the ergodic measure in the $(x,\phi)$-space.
The second term in Eq. \eqref{enrc} can be much larger than the third one, but one can check (see e.g. \cite{GRT}) that after averaging over
the ergodic measure and over the period of the billiard oscillation, the second term vanishes (this is a consequence of
the existence of the so-called Anosov-Kasuga adiabatic invariant in the ergodic case, see \cite{Jarz,GRST,GRT, Anosov, Kasuga, Brown, Lochak, MacKay}).
Corrections to the averaging due to a slow change in the billiard shape and energy were computed
in \cite{Jarz}. It follows from \cite{Jarz} that, after the averaging, the $O(uv)$-term in Eq. \eqref{enrc} effectively
acquires a small factor of order $|u|/v$ (see also \cite{GRST}). Thus, the effective change of the averaged energy per collision is of order $u^2$
(i.e. it is a certain portion of the kinetic energy of the wall).

It follows that  in the ergodic case the averaged energy of a particle grows as $\langle E (N) \rangle-E_{in}=k\frac{\bar u^{2}}{2} N$,
where $\bar u$ is the averaged wall speed (average of $|u|$). Thus, we conclude (see Eqs. (\ref{eq:eout}),(\ref{5})) that in the small hole limit
\begin{equation}
E_{out}-E_{in}=k\frac{\bar u^{2}}{2}\bar{N}={k}\frac{V}{Lh} \frac{\bar u^{2}}{2}
\label{eq:egainerg}\end{equation}
for some coefficient $k$ that may depend on the billiard shape and on the details of the protocol
of the billiard wall oscillations. By plugging this result into Eq. \eqref{1},
we obtain that the energy gain rate in the ergodic case is positive,
independent of the hole size $h$, and is given by
\begin{equation}
G(E_{in})=k v_{in}\frac{V}{L} \frac{\bar u^{2}}{2},\label{7}
\end{equation}
i.e. it is proportional to the kinetic energy of the billiard wall,
to the volume of the billiard, and inverse proportional to the time
$L/v_{in}$ the gas particle with the speed $v_{in}$ needs to traverse
the billiard once.

Next, we investigate the case of a multi-component accelerator.
In this case the ergodicity of the fast motion is broken, so the $O(uv)$ term in Eq. (\ref{enrc}) does not average out.
This means that the random walk (\ref{eq:reflection}) in the velocity space acquires a non-vanishing bias,
so the particle speed is linear in $N$ and its energy is quadratic in $N$.
A more precise
description of this process is done based on the theory developed in \cite{GRST,GRT}.
We note that the time between two consecutive collisions 
tends to zero as the particle speed grows,
$t_{_{N+1}}-t_{_N} \sim L/v_{_N}$ and  it follows from Eq. (\ref{enrc}) that
$\frac{\Delta E}{\Delta t}=\frac{E_{_{N+1}}-E_{_N}}{t_{_{N+1}}-t_{_N}}\sim  L u_{_N} \cos\phi_{_N} E_{_N}$,
i.e. in the non-ergodic case the energy changes exponentially with time, with a certain random rate.
 On a longer time scale this process can be modeled by a
multiplicative random walk (see \cite{GRST,GRT,TD}):
\begin{equation}
E_{n+1}=\xi_{n}^{2}E_{n}, \qquad v_{n+1}=\xi_{n}v_{n}, \label{8}
\end{equation}
where $E_{n}=\frac{v_n^2}{2}$ is the kinetic energy after $n$ periods of the billiard
oscillations, and $\xi_{n}$ is the sequence of independent, identically
distributed random variables, independent of the initial energy. Importantly, it is shown in \cite{GRT,TD}
that this random walk cannot be decelerating and, typically,
\begin{equation}
{\mathbb{E}}\ln\xi_{n}>0, \qquad {\mathbb{E}}\xi_{n}>1,\quad{\mathbb{E}}\xi_{n}^{2}>1.  \label{9}
\end{equation}

If we ignore  details of particle behavior on the time scales below
the period $T$ of billiard oscillations, we can infer from Eq. \eqref{8}
the following description for the behavior of the averaged energy and speed gain at time $t$:
\begin{equation}
\langle E(t)\rangle =E_{in}e^{\mu t},\qquad \langle v(t)\rangle=v_{in}e^{\lambda t},\label{11}
\end{equation}
\begin{equation}\label{vs1s2} \begin{array}{l}
\langle v(s_{1})v(s_{2}) \rangle = \langle v^{2}(s_{1}) \rangle \;\cdot\;\langle v(s_{2})/v(s_{1}) \rangle=\\ \\ \displaystyle
\!\!\!\!\!\!= \langle v^{2}(s_{1}) \rangle e^{\lambda(s_{2}-s_{1})}\!\!=
2E_{in} e^{\mu s_{1}}e^{\lambda(s_{2}-s_{1})}\quad(s_{2}\geq s_{1}),
\end{array}\end{equation}
where $\mu=\frac{1}{T}\ln{\mathbb{E}}\xi^{2}\;>\;2\lambda=\frac{2}{T}\ln{\mathbb{E}}\xi\;>\;0$.

Note that the number of collisions up to time $t$ can be related to the particle speed via
$L N(t)\sim \int_{0}^{t}v(s)ds$.
So, the averaged number of collisions up to time $t$ is given by
$$ \langle N(t)\rangle = k_1 v_{in} \frac{(e^{\lambda t}-1)}{\lambda L}$$
(see Eq. (\ref{11})) and, by Eq. (\ref{vs1s2}),
\begin{eqnarray*}
\langle N^{2}(t)\rangle  & = & \frac{k_2}{{L}^{2}}\int_{0}^{t}\int_{0}^{t}\langle v(s_{1})v{(}s_{2})\rangle ds_{1}ds_{2}\\
 & = & \frac{2k_2}{{L}^{2}}\int_{0}^{t}\int_{s_{1}}^{t}\langle v(s_{1})v(s_{2})\rangle ds_{2}ds_{1} \\
& = &4k_2 \frac{E_{in}}{L^{2}}\int_{0}^{t}\int_{s_{1}}^{t}e^{\mu s_{1}}e^{\lambda(s_{2}-s_{1})}ds_{2}ds_{1} \\
& = & 4k_2 \frac{E_{in}}{(\mu-\lambda) L^{2}}\left[\frac{e^{\mu t}-1}{\mu}-\frac{e^{\lambda t}-1}{\lambda}\right],
\end{eqnarray*}
where $k_{1,2}$ are some coefficients of order $1$.

Rearranging the expressions for $\langle N\rangle$ and $\langle N^2\rangle$, we get
$$\langle E(t)\rangle -E_{in} = 2k_1\mu v_{in} L \langle N(t)\rangle + 4k_2 \mu^2 L^2 \left(1-\frac{\lambda}{\mu}\right) \langle N(t)^2\rangle,$$
which, after averaging over the time $t$ the particle resides in the billiard, gives
\begin{equation}\!\!\!
E_{out}\!-E_{in}\!=\mu \frac{L^2}{T} \!\left[2 k_{1}\frac{v_{in}T}{L} \bar{N}\! + 4 k_2\! \left(\!1\!-\frac{\lambda}{\mu}\!\right) \mu T \overline{N^{2}}\right]\!.
\label{eq:egainmc0}\end{equation}
The rate $\mu$ in this formula is a well defined quantity determined by a one period run of the accelerator,
and an analytic expression for $\mu$ is also available in many cases \cite{GRT,STR,GRST,Shah,GRT14,Benjamin,TD}. However, to compare the energy
gain with that given by Eq. \eqref{eq:egainerg} in the ergodic case, we need to relate the rate $\mu$ with $\frac{\bar u^2}{2}$,
the kinetic energy of the wall.

By Eq. \eqref{9} the exponential growth rate $\mu$ is always non-negative. The minimal value $\mu=0$ is achieved
when the distribution of particles in the billiard remains uniform during the period of billiard oscillations,
like in the case of ergodic billiard. Hence, when the deviation from the ergodic behavior is small, the rate $\mu$ is small and
behaves like the square of a certain quantitative measure of this deviation. The violation of ergodicity in the exponential accelerator
is caused by changes of the phase space structure of the frozen billiard as its shape changes with time.
Therefore, we relate the deviation from ergodicity to the magnitude of the shape change over the period. As the billiard size
is changed with the mean speed $\bar u$, the dimensionless parameter estimating the ergodicity violation
is $\bar uT /L$. Thus, the rate of the energy increase over the period $T$ is, at small $\bar u$, given by
\begin{equation}\label{muutl}
\mu T \propto \left(\frac{\bar u T}{L}\right)^2\!.
\end{equation}
One can also extract this relation from the formulas for $\mu$  for various cases of multi-component accelerators (e.g. from \cite{GRT,STR,GRST,GRT14,Benjamin}).
Notice that even when $\bar u$ is not small, the ratio $\bar u T/L$ remains bounded and so does $\mu T$ (we
assume everywhere that the typical length scale of the billiard does not change hugely along the cycle; otherwise pathological behaviors may arise).
So, Eq. \eqref{muutl} can be used in this case as well; it would then simply mean that the quantities on both side of the relation are of order $1$.

Plugging Eq.~(\ref{muutl}) into Eq.~\eqref{eq:egainmc0}, we find
\begin{equation}
E_{out}-E_{in}=\frac{{\bar u}^2}{2} \left[k_{1}\frac{v_{in}T}{L}+k_{2} \mu T \overline{N^{2}}\right],
\label{eq:egainmc}\end{equation}
or, in the case $\bar u T/L \ll 1$,
\begin{equation}
E_{out}-E_{in}=\frac{{\bar u}^2}{2}\left[k_{1}\frac{v_{in}T}{L} \bar{N}+k_{2} \left(\frac{\bar{u}T}{L}\right)^{2} \overline{N^{2}}\right],
\label{eq:egainmc1}\end{equation}
where the new coefficients $k_{1,2}$ depend on the shape of the billiard and the protocol of the wall oscillation.

The term $v_{in}T/L$  is proportional to the number of collisions per period. In our setting this number is assumed to be large. It is also important for
the validity of the exponential growth model that the particle is initially fast (i.e. $v_{in}\gg \bar u$) yet it 
remains in the billiard at least for one period. Therefore, Eqs.
(\ref{eq:egainmc0}), (\ref{eq:egainmc}), and (\ref{eq:egainmc1}) are valid under the assumption
$$1 \ll \frac{v_{in}T}{L}\lesssim \bar{N};$$
in particular \(v_{in} \lesssim \frac{V}{T h}\). When this condition is violated, our theory is not applicable.

Comparing Eqs. (\ref{eq:egainmc}), (\ref{eq:egainmc1}) with Eq. \eqref{eq:egainerg} we see that the energy gain in the multi-component accelerator is much larger
than in the ergodic case. Even if the exponential growth rate $\mu$ is very small, the coefficient of \(\bar u^2 \bar{N}\) 
in Eq.~(\ref{eq:egainmc}) is large whereas the corresponding
coefficient in  Eq.~\eqref{eq:egainerg} is simply a constant. With the increase of $\mu$ the quadratic in \(\bar{N}\) term becomes dominant in
Eqs.~(\ref{eq:egainmc}), (\ref{eq:egainmc1}) and provides the main contribution to the energy gain.

Using Eq. \eqref{5} for $\bar N$ and $\overline{N^2}$, we finally find the
energy production rate of Eq.~\eqref{1} for the multi-component case:
\begin{equation}
G(E_{in})=v_{in}\frac{V}{L}\frac{\bar u^{2}}{2}\left[k_{1}\frac{v_{in}T}{L}+k_{2} \mu T \frac{V}{Lh}\right].\label{18}
\end{equation}
Clearly, the gain rate $G$ can be made much larger than the gain in the ergodic case by diminishing the hole size or by increasing the incoming velocity.

\section{Simulations \label{sec:Simulation-Results}}

There are two distinct predicted dependencies of the energy
gain on the hole size and on \(v_{in }\) for the two types of leaky accelerators.  To examine these predictions we consider two classes of  leaky accelerators where each of these is considered with two sets of parameters - one corresponding to an ergodic case and the other to a multi-component case.

\textbf{Dispersing  accelerators} (Figs.~\ref{fig:fig1}a,b): at \(t=0 \) a vertical bar is inserted at a position \(x_{b}\), to the double Sinai billiard
(a rectangle with two discs). Then the bar moves to the right with a constant velocity  \(u\) till time \(\tau\) and then the bar is removed. The cycle restarts at time \(T\).
We consider two cases: 1a) ergodic case where the bar only partially blocks the rectangle, covering  90\% of the rectangle length (this case is called ``Sinai'' in all Figures); 1b) multi-component case where the bar completely divides the rectangle into two parts (hereafter ``divided Sinai"). Notice that each component of the frozen billiards is ergodic and mixing \cite{Sinai}. These accelerators exhibit exponential-in-time energy growth in the multi-component case and quadratic-in-time energy growth in the ergodic case \cite{GRST}. To examine the leaky behavior, two holes of length \(h\) are placed on the upper rectangle boundary; the holes are shifted from the disks centers to avoid fast escaping orbits, and two holes are introduced to avoid strong dependence on the billiard oscillation phase. In all simulations we use the following parameters:
the rectangle width is $a=4$, its height is $b=2$, the disks radii are $1/2$, the bar velocity is \(u= 0.1\), the bar is introduced at the position \(x_{b}=0.0915\) and removed at the time moment $\tau =1.83$, the period is $T=5.49$. The initial energy and the hole size are as indicated in the figures.

\textbf{Focusing  accelerators} (Figs.~\ref{fig:fig1}c,d):
The  mushroom is a multi-component system having an integrable component and a chaotic component  \cite{BuniMush} whereas the slanted stadium is ergodic and mixing
\cite{BuniStad}. The oscillating mushroom accelerator exhibits exponential-in-time energy growth \cite{GRT14} whereas the oscillating stadium exhibits
quadratic-in-time energy growth \cite{LRA, GRT}.

The shape of the mushroom is determined by the following four parameters:
$r$ is the radius of the cap;  $w$ is the half-width of the hole at the bottom of the cap, it
 coincides with the half-width of the stem at its highest point and $w\le r$; $\ell$ is the length of the stem;
and the angle $\theta$ describes the inclination of the stem sides. When $w=r$ the mushroom
becomes a slanted stadium.

For the purpose of numerical experiments we used the following protocols:
$r(t)=1$, $w(t) = b_0 - b_1(1 - \cos(t))$,
and $\ell(t) = a_0 - a_1 \sin(t)$,
$ a_0 =1$, $b_0 =1$. For the mushroom, we set $ a_1 = 0.5$ and $b_1 = 0.4$;
and for the  stadium $a_1 = 0.5$, $b_1= 0$.
The hole is located at the bottom of the stem with the center displaced by $0.01$
from the center of the stem. In all experiments $\theta = 0.1111$.

In each numerical experiment 2000 particles are injected at randomly chosen times during the period \([0,T] \) into the billiard through the holes,
with random position in the hole and entering angle.
Each particle moves inside the billiard undergoing elastic collisions with the boundary till it
exits by colliding with the hole.
The exit time, the number of collisions till exit, and the exit speed are recorded.

Figure~\ref{fig:fig2}a demonstrates that the average number of collisions at exit, $\overline N$, scales linearly with $1\big/h$ for both the multi-component and the
ergodic cases. Figure \ref{fig:fig2}b shows that $\overline N$ does not depend on the initial energy, again for both cases. Figure \ref{fig:fig2}c shows
the geometric distribution of the exit collisions for the mushroom. These results support the assumptions made in Eq.~\eqref{5}.

Figure~\ref{fig:fig3} shows the dependence of  $G/v_{in}=(E_{out}-E_{in}) h$ on $1\big/h$ for the four billiard types. Figure \ref{fig:fig3}a shows
that the net energy flow increases linearly with $1\big/h$ for billiards with exponential acceleration (non-ergodic case) as predicted in Eq.~\eqref{18}.
The inset shows that for sufficiently small holes the flow is essentially independent of the hole size for the ergodic billiards as predicted by Eq.~\eqref{7}.

Figure~\ref{fig:fig4} shows that the average energy gain, $E_{out}-E_{in}$,  grows linearly with $v_{in}$ for the multi-component cases as predicted by
Eq.~(\ref{eq:egainmc}). The inset shows that this difference also grows with $v_{in}$  for the ergodic case (at a slow pace).
To the leading order in \(h\) one should not see such growth according to Eq.~(\ref{eq:egainerg}).
We explain the effect by the order $h$ difference between
the statistics of the closed and leaky accelerators. For example, averaging the second term of Eq.~(\ref{enrc})
over a boundary with the order $h$ hole produces corrections of order \(h \bar u v \),
which cause an order \(h\) bias in the random walk of the velocity.
Figure~\ref{fig:fig5} shows the mean bias $\left<v_{out}-v_{in}\right>\big/\left<N\right>$, which is indeed present and does not vanish
in the limit of large initial speed, both in the ergodic and multi-component case. However, the bias is much smaller
in the ergodic case, confirming our conjecture that it is of order \(h\).
The dependence of the energy gain on \(\bar u\) appears to be a more delicate numerical issue and will be further investigated elsewhere.

\section{Discussion and Conclusion \label{sec:Discussion-and-Conclusion}}

A priory, a faster accelerator does not automatically imply higher energy gain
as faster particles escape earlier from the accelerator and
the net energy gain reflects the balance between the averaged escape time and the energy gained by then.
In this paper we find that  in a  billiard
the escape is mainly determined by the average number of collisions,
and from this observation we are able to find the balance
between these  two factors for two different types of accelerators. Our results conclusively show that multi-component leaky accelerators produce higher net energy flow than
their ergodic counterparts. Moreover, we show that the difference increases
significantly when the hole size decreases.

One may worry that decreasing the hole size will lead to very long residence times.
However, for the multi-component case the residence time increases only logarithmically with the hole size.
Indeed, the number of collisions is related to the residence time via
$L N(t)\sim \int_{0}^{t}v(s)ds$, hence,
in the multi-component case the averaged residence time is proportional to \(\log \bar N\sim \log h\), while in the ergodic case it is proportional to
\(\sqrt{\bar N} \sim h^{-1/2}\). Hence, we expect that taking the small hole limit for gaining energy may become practical in the case of multi-component
leaky accelerators: they produce high gain at a fast pace.

\acknowledgements
VG research is supported by the EPSRC grant EP/J003948/1.
VRK, The Estrin Family Chair of Computer Science
and Applied Mathematics, thanks the ISF (grant 321/12) for its support.
D.T. is supported by Grant No. 14-41-00044 of RSF (Russia) and by the Royal Society grant IE141468.

\end{document}